\documentclass[aps,prb,a4paper,10pt,twocolumn,showpacs,floatfix,superscriptaddress,preprintnumbers,longbibliography]{revtex4-2}
\usepackage{float}
\usepackage{dcolumn,graphicx,color,booktabs,microtype,afterpage}
\usepackage{amssymb}
\usepackage{amsmath}
\usepackage[charter,greekuppercase=italicized]{mathdesign}
\usepackage{sidecap}
\usepackage[mathlines]{lineno}
\usepackage{gensymb}

\graphicspath{{./}{figure/}}
\renewcommand{\tablename}{Table}
\makeatletter\renewcommand{\fnum@figure}[1]{\figurename~\thefigure.~}\makeatother
\makeatletter\renewcommand{\fnum@table}[1]{\tablename~\thetable.}\makeatother

\newcount\hh \newcount\mm
\hh=\time \divide\hh by 60
\mm=\hh \multiply\mm by 60 \mm=-\mm
\advance\mm by \time
\def\now{\number\hh:\ifnum\mm<10{}0\fi\number\mm}

\usepackage[colorlinks,plainpages=false,linkcolor=blue,urlcolor=blue,citecolor=blue,pdfpagemode=UseNone,pdfstartview=FitBH]{hyperref}
\usepackage{nicefrac}

\newcommand{\tcr}[1]{\textcolor{black}{#1}}

\begin{document}
	\makeatletter\renewcommand{\ps@plain}{%
		\def\@evenhead{\hfill\itshape\rightmark}%
		\def\@oddhead{\itshape\leftmark\hfill}%
		\renewcommand{\@evenfoot}{\hfill\small{--~\thepage~--}\hfill}%
		\renewcommand{\@oddfoot}{\hfill\small{--~\thepage~--}\hfill}%
	}\makeatother\pagestyle{plain}
	
	\preprint{\textit{Preprint: \today, \now}} %For internal use only, do not distribute.}}
%\linenumbers

%\title{Nodeless superconductivity with a preserved time-reversal symmetry in molybdenum carbide superconductor}
%\title{Nodeless superconductivity with preserved time-reversal symmetry in molybdenum carbide}
%\title{Superconductivity and topological aspects of the rocksalt carbides NbC and TaC}
\title{Nodeless superconductivity and topological nodal states in molybdenum carbide}

%	Mo$_2$C: a {\textmu}SR study}

%\title{Superconductivity and electronic band structure of Mo$_2$C: a {\textmu}SR study}
%\title{Superconductivity and electronic band structure of Mo$_2$C}
%and possible topological superconductivity in NbC}}
%
\author{Tian\ Shang}\thanks{These authors contributed equally}\email[Corresponding authors:\\]{ tshang@phy.ecnu.edu.cn}
\affiliation{Key Laboratory of Polar Materials and Devices (MOE), School of Physics and Electronic Science, East China Normal University, Shanghai 200241, China}
\author{Yuting\ Wang}\thanks{These authors contributed equally}
\affiliation{Co-Innovation Center for New Energetic Materials, Southwest University of Science and Technology, Mianyang, 621010, People's Republic of China}
\affiliation{School of Science, Southwest University of Science and Technology, Mianyang 621010, P. R. China}
\author{Bochen\ Yu}
\affiliation{Key Laboratory of Polar Materials and Devices (MOE), School of Physics and Electronic Science, East China Normal University, Shanghai 200241, China}
\author{Keqi\ Xia}
\affiliation{Key Laboratory of Polar Materials and Devices (MOE), School of Physics and Electronic Science, East China Normal University, Shanghai 200241, China}
\author{Darek J.\ Gawryluk}%\thanks{On leave from Institute of Physics, Polish Academy of Sciences, Aleja Lotnikow 32/46, PL-02-668 Warsaw, Poland.}
\affiliation{Center for Neutron and Muon Sciences, Paul Scherrer Institut, CH-5232 Villigen PSI, Switzerland}
\author{Yang\ Xu}
\affiliation{Key Laboratory of Polar Materials and Devices (MOE), School of Physics and Electronic Science, East China Normal University, Shanghai 200241, China}
\author{Qingfeng\ Zhan}
\affiliation{Key Laboratory of Polar Materials and Devices (MOE), School of Physics and Electronic Science, East China Normal University, Shanghai 200241, China}
\author{Jianzhou\ Zhao} \email[Corresponding authors:\\]{jzzhao@swust.edu.cn}
\affiliation{Co-Innovation Center for New Energetic Materials, Southwest University of Science and Technology, Mianyang, 621010, People's Republic of China} 
%
%
%\author{Ming Shi}
%\affiliation{Swiss Light Source, Paul Scherrer Institut, Villigen CH-5232, Switzerland}
%
%
\author{Toni Shiroka}%\email[Corresponding authors:\\]{tshiroka@phys.ethz.ch}
\affiliation{Center for Neutron and Muon Sciences, Paul Scherrer Institut, CH-5232 Villigen PSI, Switzerland}
\affiliation{Laboratorium f\"ur Festk\"orperphysik, ETH Z\"urich, CH-8093 Z\"urich, Switzerland}
\begin{abstract}
The orthorhombic molybdenum carbide superconductor with $T_c$ = 3.2\,K was investigated by muon-spin rotation and relaxation ({\textmu}SR) measurements and by 
first-principle calculations. The low-temperature superfluid density, determined by transverse-field {\textmu}SR, suggests a fully-gapped superconducting state in Mo$_2$C, with a zero-temperature gap $\Delta_0$ = 0.44\,meV and a magnetic penetration depth $\lambda_0$ = 291\,nm. The time-reversal symmetry is preserved in the superconducting state, as confirmed by the absence of an additional muon-spin relaxation in the zero-field {\textmu}SR spectra. Band-structure calculations indicate that the density of states at the Fermi level is dominated by the
Mo $4d$-orbitals, which are marginally
hybridized with the C $2p$-orbitals over a wide energy range.
The symmetry analysis confirms that, in the absence of spin-orbit coupling (SOC), Mo$_2$C hosts twofold-degenerate nodal surfaces and fourfold-degenerate nodal lines. 
When considering SOC, the fourfold-degenerate nodal lines cross the Fermi level and contribute to the electronic properties.
Our results suggest that, similarly to other phases of carbides, also the orthorhombic
transition-metal carbides host topological nodal states and may be potential candidates for future studies of topological superconductivity.
\end{abstract}

%\keywords{Unconventional supeconductivity, muon spin rotation/relaxation, nuclear magnetic resonance}

\maketitle\enlargethispage{3pt}

\vspace{-5pt}
\section{\label{sec:Introduction}Introduction}\enlargethispage{8pt}
The possibilities offered by topological superconductors, ranging from hosting Majorana fermion quasiparticles to potential applications in topological quantum computing~\cite{Qi2011,Kitaev2001,Liang2011,Sato2017}, have stimulated the researchers to explore different routes to realize them. 
The most obvious approach consists in the introduction of extra carriers into a topological insulator to achieve superconductivity (SC). This route has been frequently attempted
in the copper- or strontium intercalated Bi$_2$Se$_3$ topological insulator~\cite{Hor2010,Sasaki2011,Liu2015,Li2019}. Another approach utilizes the proximity effect between a conventional $s$-wave superconductor and a topological insulator or semiconductor~\cite{Li2019,Lutchyn2018}. 
The surface states of  a topological insulator can lead to a two-dimensional superconducting state with a $p+ip$ pairing at the interfaces, known to support Majorana bound states at the vortices~\cite{Liang2008}. For instance, evidence of topological SC has been reported in
NbSe$_2$/Bi$_2$(Se,Te)$_3$ heterostructures~\cite{Xu2014,Peng2014}, where NbSe$_2$ represents a typical fully-gapped superconductor.

Despite continued efforts to identify topological SC in accordance with
the aforementioned approaches,
the intricacy of heterostructure fabrication, the rarity of suitable
topological insulators, and the inhomogeneity or disorder
effects induced by carrier doping, have
considerably constrained the investigation and potential applications of topological SC.
A more attractive way to achieve them is to combine
superconductivity and a nontrivial electronic band structure in the same material.
Clearly, it is of fundamental interest to be able to identify such new superconductors with nontrivial band topology, but
with a simple composition. 
For example, topologically protected surface states have been found in superconducting CsV$_3$Sb$_5$~\cite{Hu2022}, $\beta$-PdBi$_2$~\cite{Sakano2015}, and PbTaSe$_2$~\cite{Guan2016}, 
all of which are good platforms for studying topological SC. 

In this respect, the binary transition-metal carbides (TMCs)
represent another promising family of materials.
TMCs exhibit essentially four different solid phases, which include the $\alpha$ ($Fm\overline{3}m$ No.\,225)-,
$\beta$ ($Pbcn$, No.\,60)-, $\gamma$ ($P\overline{6}m2$, No.\,187)-, and
$\eta$ ($P6_3/mmc$, No.\,194)-phase~\cite{Toth1971}. The $\gamma$-phase is noncentrosymmetric, while the other three are centrosymmetric. Due to the lack of space inversion, the $\gamma$-phase TMCs exhibit exotic topological features. The unconventional three-component fermions with surface Fermi arcs were experimentally observed in $\gamma$-phase WC~\cite{Ma2018}.
By applying external pressure, the topological semimetal MoP (isostructural to WC) becomes a superconductor, whose $T_c$ rises up to 4\,K (above 90\,GPa)~\cite{Chi2018}, thus representing a candidate topological superconductor. Unfortunately, no SC has been observed in WC yet, but the $\gamma$-phase MoC (similar to WC and also not superconducting) was predicted to be a topological nodal-line semimetal with drumhead surface states~\cite{Huang2018}. The $\alpha$-phase TMCs show a relatively high $T_c$ value and some of them
were also predicted to exhibit nontrivial band topologies~\cite{Huang2018,Zhan2019,Isaev2007,Tutuncu2012,Kavitha2016,Szymanski2019,Shang2020,Willens1967,Williams1971,Toth1971}.  
For example, NbC and TaC are fully-gapped superconductors with $T_c = 11.5$ and 10.3\,K, respectively~\cite{Shang2020}. 
At the same time, theoretical calculations suggest that
$\alpha$-phase TMCs are  nodal line semimetals in the absence of spin-orbit coupling (SOC)~\cite{Shang2020,Huang2018}.

As for the molybdenum carbides, although their SC was already reported in the 1970s~\cite{Willens1967}, their physical properties have been overlooked due to difficulties in synthesizing clean samples. Only recently, the $\alpha$-phase MoC$_x$ ($x < 1$) and $\eta$-phase Mo$_3$C$_2$ (with $T_c$ = 14.3 and 8.5\,K) could be synthesized under high-temperature and high-pressure conditions (1700\,$^\circ$C, 6--17\,GPa)~\cite{Yamaura2006,Sathish2012,Sathish2014} and their superconducting properties studied via different techniques. 
To date, the electronic properties of the other phases
of molybdenum carbides (e.g., the $\beta$-phase) remain mostly unexplored.  

In this paper, we report on the superconducting properties of the
$\beta$-phase Mo$_2$C, investigated via magnetization- and muon-spin relaxation and rotation ({\textmu}SR)
measurements. In addition, we also present numerical density-functional-theory (DFT) band-structure calculations. 
We find that Mo$_2$C exhibits a fully-gapped superconducting state,
while its electronic band structure suggests that 
it hosts twofold-degenerate nodal surfaces and fourfold-degenerate nodal lines. 
Therefore, the $\beta$-phase %orthorhombic 
TMCs (of which Mo$_2$C is a typical example)
may be potential candidates for future studies of topological SC,
similar to the other TMC phases.

\section{Experimental and numerical methods\label{sec:details}}\enlargethispage{8pt}
First, we tried to synthesize the $\beta$-phase Mo$_2$C by arc melting
Mo slugs (99.95\%, Alfa Aesar) and C rods (99.999+\%, ChemPUR).
Similarly to previous studies~\cite{Mao2019}, the obtained
polycrystalline samples showed a mixture of different phases, both before and after the annealing. Akin to the $\alpha$-phase, the $\beta$-phase Mo$_2$C can be synthesized also under high-temperature and high-pressure conditions (1500–2300\,K, 5\,GPa)~\cite{Ge2021}. 
However, the resulting Mo$_2$C samples have a rather low superconducting volume fraction. 
Because of these issues, all our measurements were performed
on high-purity Mo$_2$C powders (99.5\%) produced by
Alfa Aesar.
For the {\textmu}SR investigation,
the powders were pressed into pellets, while for the magnetization
measurements, performed on a 7-T Quantum Design magnetic property
measurement system, loose powders were used. 
Room-temperature x-ray powder diffraction (XRD) 
was performed on a Bruker D8 diffractometer using Cu K$\alpha$ radiation. 
The {\textmu}SR measurements were carried out at the multipurpose surface-muon spectrometer (Dolly) at the $\pi$E1 beamline of the Swiss muon source at Paul Scherrer Institut (PSI), Villigen, Switzerland. 
The Mo$_2$C pellets were mounted on 25-{\textmu}m thick copper foil to
cover an area 6--8\,mm in diameter. The {\textmu}SR spectra comprised
both transverse-field (TF) and zero-field (ZF) 
measurements, performed upon heating the sample.
The {\textmu}SR spectra were analyzed by means of the \texttt{musrfit} software package~\cite{Suter2012}.

The phonon spectrum and the electronic band structure of Mo$_2$C
were calculated via DFT, within the generalized gradient approximation (GGA) of Perdew-Burke-Ernzerhof (PBE) realization~\cite{Perdew:1996iq}, as implemented in the Vienna \emph{ab initio} Simulation Package (VASP)~\cite{Kresse:1996kl,Kresse:1996vk}. The projector augmented wave (PAW) pseudopotentials were adopted for the calculation~\cite{Kresse:1999wc,Blochl:1994zz}. Electrons belonging to the outer atomic configuration were treated as valence electrons, here
corresponding to 6 electrons in Mo ($4d^55s^1$) and 4 electrons in C ($2s^22p^2$). The kinetic energy cutoff was fixed to 400\,eV. 
For the three different crystal structures of Mo$_2$C,  the atomic
positions and the lattice constants were fully relaxed for the
calculations of the phonon dispersion spectrum. 
The force convergence criterion was set to 1\,meV.
For the structure optimization calculations, Monkhorst-Pack grids
of $16 \times 16 \times 10$, $14 \times 11 \times 13$, and
$19 \times 19 \times 21$ $k$-points were used for the space groups
$P6_3/mmc$, $Pbcn$, and $P\bar{3}1m$, respectively.
To obtain the force constants and phonon spectra, we used the
density functional perturbation theory (DFPT)
in combination with the Phonopy package~\cite{gonze_dynamical_1997,phonopy-phono3py-JPCM,phonopy-phono3py-JPSJ}.
A supercell of $2\times 2\times 2$ was adopted for the calculation of force constants.
To calculate the phonon spectrum, the Brillouin zone integration
was performed on a $\Gamma$-centered mesh of $10 \times 10 \times 7$,
$7 \times 5 \times 6$, and $6 \times 6 \times 7$ $k$-points for
the space groups $P6_3/mmc$, $Pbcn$, and $P\bar{3}1m$, respectively. 
In the $P6_3/mmc$ case, considering that only half of the 2$a$ sites is occupied by C atoms, we simplified the structure such that C atoms are fully occupied only at the corners of the unit cell.
The spin–orbit coupling (SOC) was fully considered in our calculation. 
After optimizing the parameters, the electronic- and phononic band structures, as well as the density of states (DOS)
were calculated.

\section{\label{sec:results }Results and discussion}\enlargethispage{8pt} 
\subsection{\label{ssec:structure} Crystal structure}
\begin{figure}[t]
	\centering 
	\includegraphics[width=0.48\textwidth,angle=0]{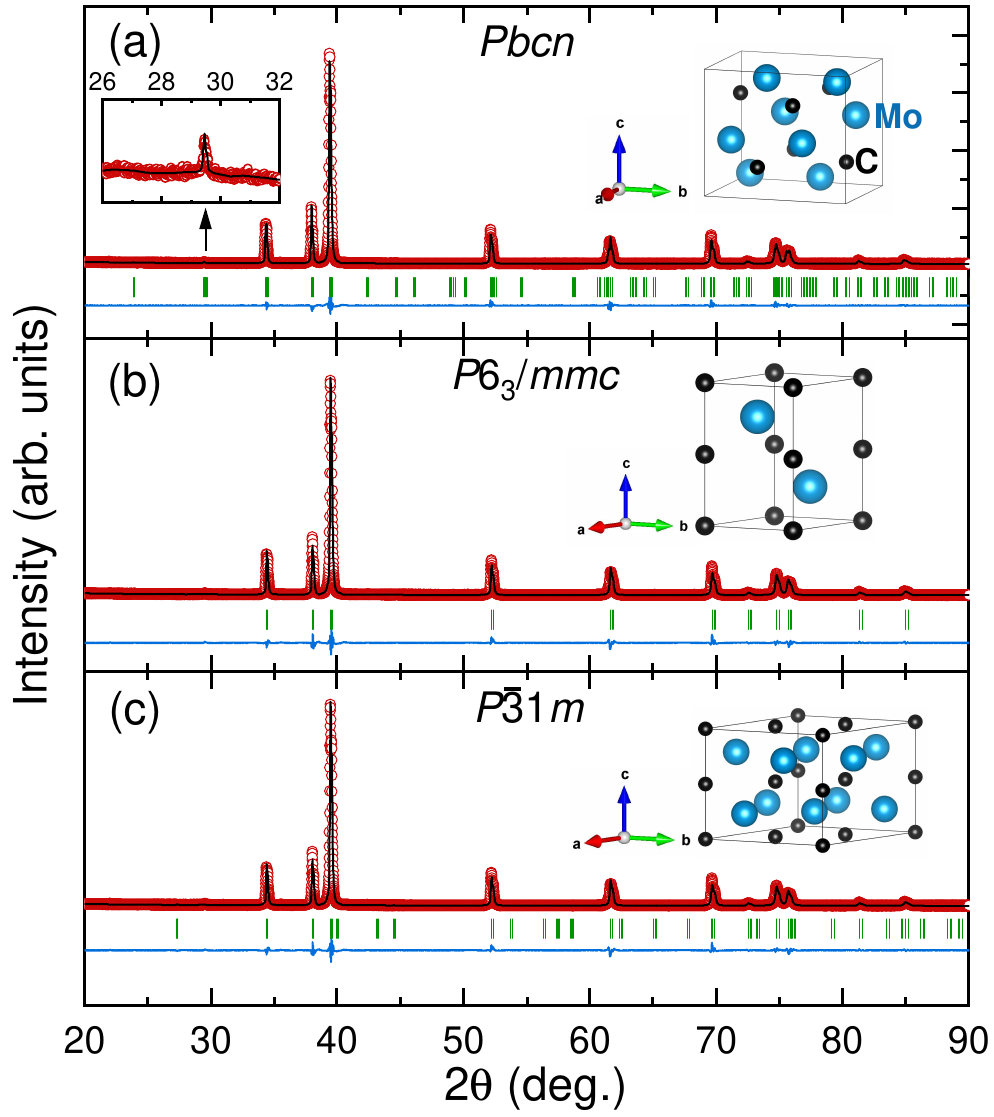} 
	\caption{\label{fig:XRD}Room-temperature x-ray powder diffraction 
		patterns and Rietveld refinements for Mo$_2$C powders using
		different space groups: $Pbcn$ (a), $P6_3/mmc$ (b), and
		$P\bar{3}m$ (c). The left inset in (a) shows an enlarged
		plot of intensity for 2$\theta$ between 26 and 30$^\circ$.
		The open red circles and the solid black lines represent
		the experimental patterns and the refinement profiles,
		respectively. The blue lines at the bottom
		show the residuals, i.e., the difference between the calculated 
		and the experimental data. The vertical bars mark the calculated Bragg-peak 
		positions. The unit-cell crystal structures are shown
		in the right insets of each panel. Here, the blue
		and black spheres represent the Mo and C atoms, respectively.}
\end{figure}

The phase purity and the crystal structure of Mo$_2$C powders were
checked via XRD measurements at room temperature (see Fig.~\ref{fig:XRD}).
Unlike the arc-melted Mo$_2$C~\cite{Mao2019}, the purchased Mo$_2$C
powders show a clean phase. Several phases of molybdenum carbides
have been reported, which exhibit cubic, orthorhombic, hexagonal,
and trigonal structures~\cite{Sathish2012,Sathish2014,Ge2021,Rudy1968,Epicier1988}.
In our case, the XRD pattern of Mo$_2$C, was analyzed by means of
the FullProf Rietveld-analysis suite~\cite{Carvajal1993} to find that
only the latter three structures, with space groups
$Pbcn$ (No.\,60, orthorhombic), $P6_3/mmc$ (No.\,194, hexagonal), and
$P\bar{3}1m$ (No.\,162, trigonal), reproduce the data reasonably well.
In the insets we depict the corresponding crystal structures,
known as $\beta$-, $\eta$-, and $\zeta$-phases, respectively.
Among these, the $\beta$-phase exhibits the best agreement with
the measured XRD pattern, here reflected in the smallest
goodness-of-fit factor (see Table~\ref{tab:lattice}). 
Moreover, both the $\eta$- and $\zeta$-phases fail to reproduce some
of the low-intensity reflections. For instance, as illustrated
in the inset of Fig.~\ref{fig:XRD}(a), while neither the $\eta$- nor
the $\zeta$-phases admit a reflection at $2\theta \approx 30^\circ$,
the $\beta$-phase captures this reflection quite well. 
In conclusion, the Rietveld refinements suggest that the investigated Mo$_2$C powders
adopt an orthorhombic structure with space group $Pbcn$, as
further confirmed by the calculated phonon-dispersion spectrum (see below). Furthermore, no impurity phases could be detected, indicating a good sample quality.
The refined crystal-structure information and atomic coordinates for
all the three different phases are listed in
Tables~\ref{tab:lattice} and \ref{tab:coordinate}.

\begin{table}[tbp]
	\centering
	\caption{\label{tab:lattice} Refined lattice parameters and goodness of fits (including $R$ factors and $\chi_\mathrm{r}^2$) for Mo$_2$C powders utilizing different space groups. The orthorhombic $\beta$-phase shows clearly the best fit parameters.}
	\begin{ruledtabular}	
		\vspace{3pt}
		\begin{tabular}{lccc}
			Space group      & $Pbcn$ (No.\,60)             &   $P6_3/mmc$ (No.\,194)                 &    $P\bar{3}1m$ (No.\,162)        \\ \hline
		    Structure        & orthorhombic                 &   hexagonal                             &    trigonal                      \\
                             & ($\beta$-phase)		        &   ($\eta$-phase)                           &   ($\zeta$-phase)                     \\
			$a$ (\AA{})      & 4.7332(2)                    &   3.0070(2)                             &    5.2085(2)                     \\
			$b$ (\AA{})      & 6.0292(2)                    &   3.0070(2)                             &    5.2085(2)                     \\
			$c$ (\AA{})      & 5.2055(2)                    &   4.7279(2)                             &    4.7282(2)                     \\
			$V_\mathrm{cell}$ (\AA{}$^3$)  & 148.553(9)     &   37.021(3)                             &    111.084(8)                    \\
	    	$R_\mathrm{P}$   & 3.31                         &   4.23                                  &    4.24                          \\
			$R_\mathrm{WP}$  & 4.79                         &   6.37                                  &    6.38                          \\
			$R_\mathrm{exp}$ & 1.82                         &   1.83                                  &    1.82                          \\
			$\chi_\mathrm{r}^2$         & 6.91                         &   12.1                                  &    12.2                          \\
		\end{tabular}
	\end{ruledtabular}
\end{table}

\begin{table}	
	\centering
	\caption{\label{tab:coordinate} Refined atomic coordinates and
	occupations of the three different crystal structures of
	Mo$_2$C powders. Only the orthorhombic structure, with $Pbcn$
	space group, is relevant in our case.}
	\begin{ruledtabular}	
		\vspace{3pt}
		$Pbcn$\\
		\begin{tabular}{lccccc}
			\textrm{Atom}&
			\textrm{Wyckoff}&
			\textrm{$x$}&
			\textrm{$y$}&
			\textrm{$z$}&
			\textrm{Occ.}\\
			\colrule %\\
		    Mo1           & 8$d$  & 0.2470(4)    & 0.1250     & 0.0798(2) 	&  1 \rule{0pt}{2.6ex} \\
			C1            & 4$c$  & 0.00000      & 0.3750(2)  & 0.2500    	&  1   \\
		\end{tabular}
		\\ \vspace{10pt}		
	     $P6_3/mmc$ \\
		\begin{tabular}{lcccccc}
			\textrm{Atom}&
			\textrm{Wyckoff}&
			\textrm{$x$}&
			\textrm{$y$}&
			\textrm{$z$}&
			\textrm{Occ.}\\ 
			\colrule %\\
		    Mo1           & 2$c$  & 0.3333       & 0.6667     & 0.2500   	&  1 \rule{0pt}{2.6ex} \\
            C1            & 2$a$  & 0.0000       & 0.0000     & 0.0000     	&  0.5   \\ 
		\end{tabular} 
		\\ \vspace{10pt}
	    $P\bar{3}1m$ \\
		\begin{tabular}{lccccl}
			\textrm{Atom}&
			\textrm{Wyckoff}&
			\textrm{$x$}&
			\textrm{$y$}&
			\textrm{$z$}&
			\textrm{Occ.}\\
			\colrule %\\
		    Mo1            & 6$k$   & 0.3333       & 0.0000       & 0.2500          & 1.00    \rule{0pt}{2.6ex} \\
			C1             & 1$a$   & 0.0000       & 0.0000       & 0.0000          & 0.97    \\
			C2             & 1$b$   & 0.0000       & 0.0000       & 0.5000          & 0.03    \\
			C3             & 2$c$   & 0.3333       & 0.6667       & 0.0000          & 0.26    \\
			C4             & 2$d$   & 0.3333       & 0.6667       & 0.5000          & 0.74    \\
		\end{tabular}	
	\end{ruledtabular}
\end{table}

\subsection{\label{ssec:sus}Magnetization measurements}

%==== figure =============================%
\begin{figure}[th]
	\centering
	\includegraphics[width=0.46\textwidth,angle=0]{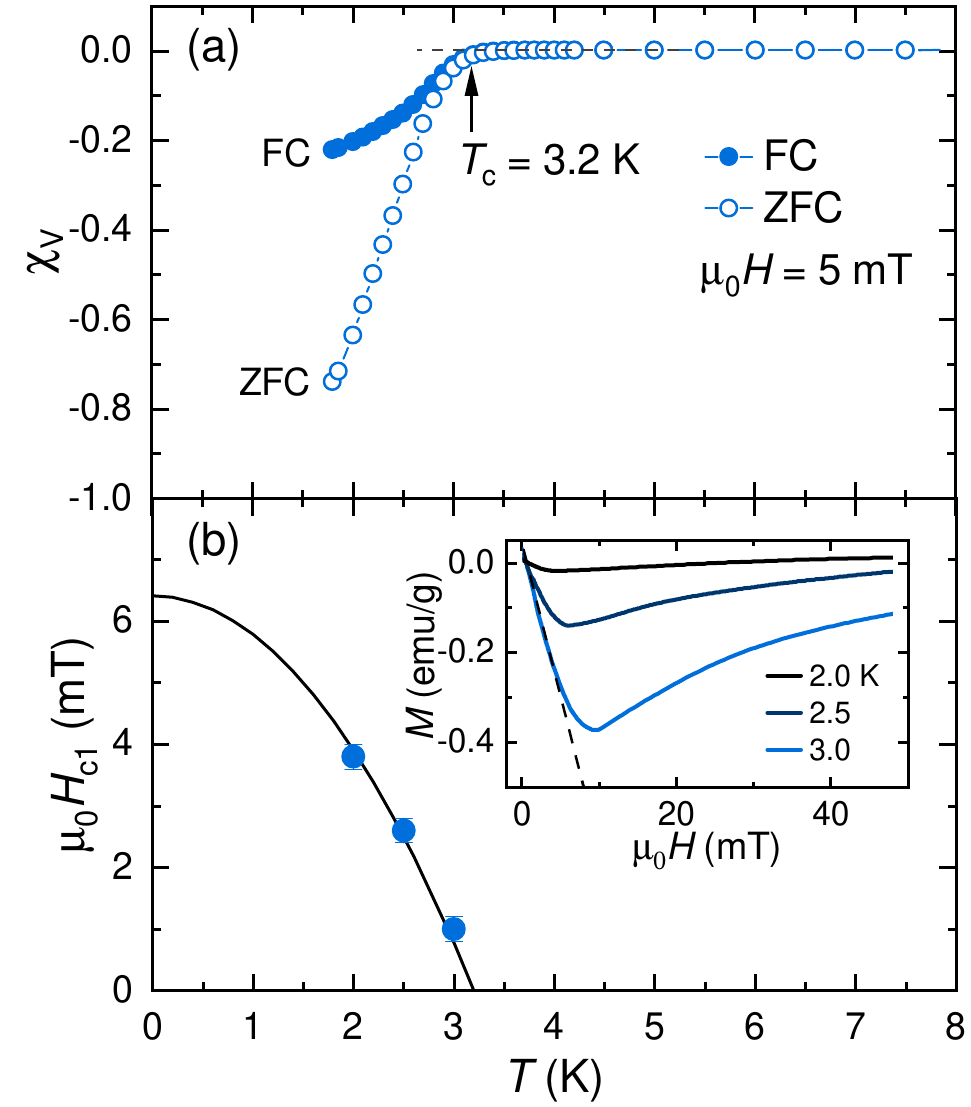}
	%	\vspace{-2ex}%
	\caption{\label{fig:Chi}(a) Magnetic susceptibility of Mo$_2$C 
		vs.\ temperature, measured in an applied field of 5\,mT using both 
		ZFC and FC protocols.
		(b) Temperature-dependent lower critical field $H_{c1}(T)$  
		for Mo$_2$C. The solid line is a fit to $H_{c1}(T) = H_{c1}(0)[1-(T/T_{c})^2]$. 
		 Inset: field\--de\-pen\-dent magnetization curves recorded
		at different temperatures below $T_c$. The lower critical field $H_{c1}$ was determined as the field value
		where $M(H)$ starts deviating from linearity (see dashed line).}
\end{figure}
%=== end figure ==========================%

We first characterized the SC of Mo$_2$C powders by magnetic susceptibility, carried out in  
a 5-mT field, using both field-cooled (FC) and zero-field-cooled (ZFC) protocols. As indicated by the arrow in Fig.~\ref{fig:Chi}(a), a clear diamagnetic signal appears
below the superconducting transition at $T_c = 3.2$\,K. The reduced
$T_c$ value compared to the previously reported $T_c \sim 6$\,K
\tcr{(whose SC fraction was less than 1\%)}
is most likely attributed to the varying C-content~\cite{Ge2021}.
\tcr{Such a variable $T_c$ against C-content has been previously reported
in the $\alpha$- and $\eta$-phase of molybdenum carbides~\cite{Yamaura2006,Sathish2012,Sathish2014},
where the light C atom is expected to modify significantly the electron-phonon
coupling and the phonon frequencies and, ultimately, also $T_c$.}
%Such a variable $T_c$ against the C-content has been previously reported in the $\alpha$- and $\eta$-phase molybdenum carbides, where the electron-phonon coupling and phonon frequency are mostly expected to be modified.}
 %Such a $T_c$ is slightly lower than the previously reported value~\cite{Ge2021}, which is most likely due to the variation of C-content. 
A large diamagnetic response (i.e., $\chi_\mathrm{V} \sim -0.8$ at 2\,K)
indicates a bulk SC in Mo$_2$C, as further confirmed by our TF-{\textmu}SR measurements. 
The field-dependent magnetization curves $M(H)$, collected at a few temperatures below $T_c$, are plotted in the inset of Fig.~\ref{fig:Chi}(b). The estimated lower
critical fields $H_{c1}$ as a function of temperature are summarized in Fig.~\ref{fig:Chi}(b). This yields a lower critical field $\mu_0$$H_{c1}$ = 6.4(2)\,mT for Mo$_2$C at zero temperature (see solid line).

\subsection{\label{ssec:TF_muSR}Transverse-field \texorpdfstring{{\textmu}SR}{MuSR}}
%
%
%%
%==== figure =============================%
\begin{figure}[htp]
	\centering
	\includegraphics[width=0.44\textwidth,angle= 0]{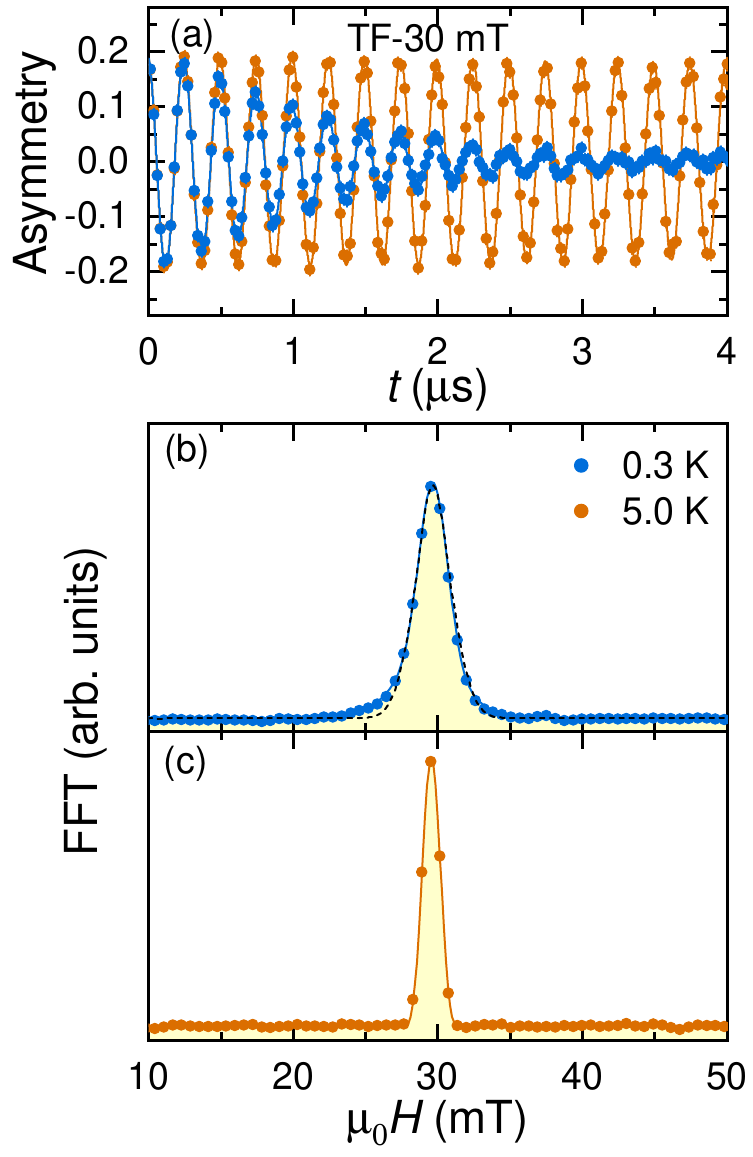}
	\caption{\label{fig:TF_muSR}(a) Representative TF-{\textmu}SR spectra of Mo$_2$C collected in 
	 its superconducting- (0.3\,K) and normal state (5\,K) in an applied magnetic field of 30\,mT. 
	Fast Fourier transforms of the relevant time spectra at  (b) 0.3\,K and (c) 5\,K. The dashed- and solid lines
	are fits to Eq.~(\ref{eq:TF_muSR}) using one and two oscillations, respectively. 
	Note the clear broadening of field distribution in the superconducting state. Fits with two oscillations show a goodness-of-fit  
	$\chi_\mathrm{r}^2$ $\sim$ 1.1, which is relatively smaller than $\chi_\mathrm{r}^2$ $\sim$ 1.7 of the one-oscillation fits.
	}
\end{figure}
%=== end figure ==========================%
%

\begin{figure}[t]
	\centering
	\includegraphics[width=0.48\textwidth,angle= 0]{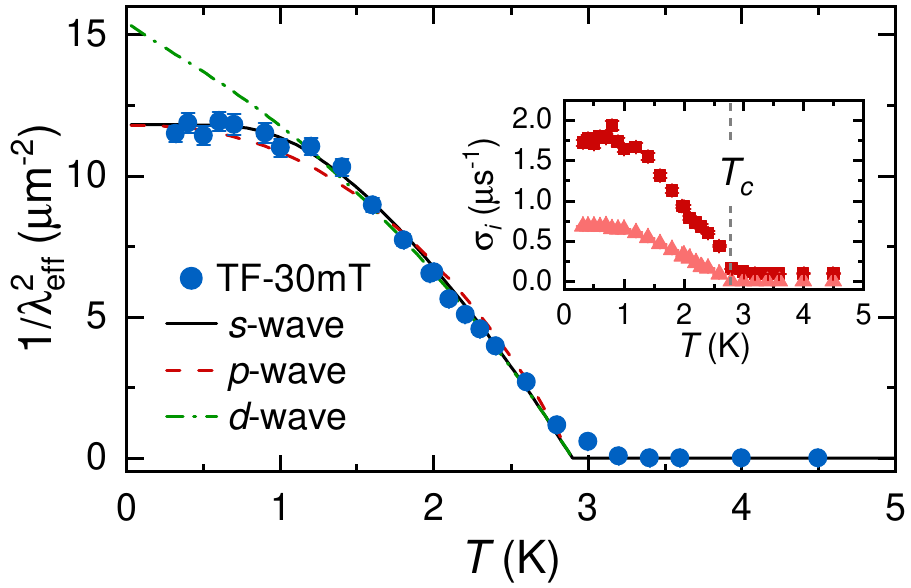}
	\caption{\label{fig:rho_sc}Temperature-dependent inverse square of the effective magnetic penetration depth of Mo$_2$C, as 
	determined from TF-{\textmu}SR measurements in an applied magnetic field 
	of 30\,mT. The solid, dashed, and dash-dotted lines represent fits to the $s$-, 
	$p$-, and $d$-wave model, \tcr{with $\chi_\mathrm{r}^2 \sim 1.9$, 5.1, and 14, respectively}. The inset shows the two muon-spin relaxation rates $\sigma_i(T)$ 
	required to fit the TF-{\textmu}SR data, \tcr{where the dashed line indicates the $T_c$ = 2.9\,K.}} 
	\end{figure}

\begin{figure}[t]
	\centering
	\includegraphics[width=0.46\textwidth,angle=0]{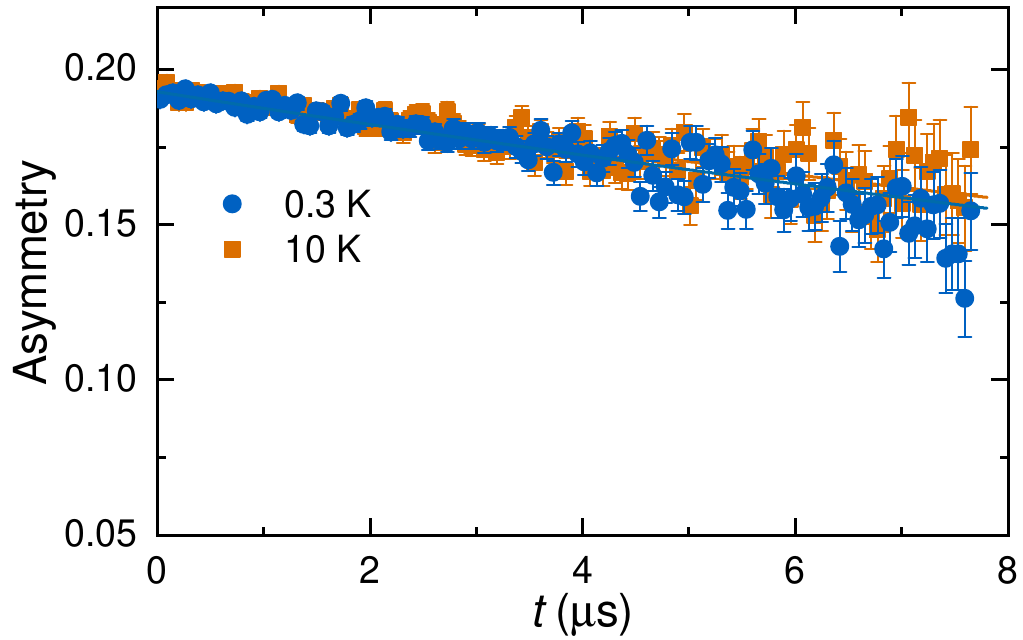}
	\vspace{-2ex}%
	\caption{\label{fig:ZF_muSR}ZF-{\textmu}SR spectra of Mo$_2$C, recorded in its su\-per\-con\-duc\-ting (0.3\,K)- and its
	normal (10\,K) states. Solid lines through the data are fits using the Kubo-Toyabe relaxation function.} 
\end{figure}

To study the gap symmetry and superconducting pairing of Mo$_2$C,
we performed systematic TF-{\textmu}SR measurements in an applied
field of 30\,mT (i.e., much higher than $H_\mathrm{c1}$) at various temperatures. 
In a TF-{\textmu}SR measurement, the magnetic field is applied perpendicular to the muon-spin direction, leading to the precession of the muon spin. 
By performing TF-{\textmu}SR, one can quantify the additional field-distribution broadening due to the flux-line lattice (FLL) and, thus, determine the superfluid density in type-II superconductors. 
Figure~\ref{fig:TF_muSR}(a) plots two representative superconducting- and  normal-state TF-{\textmu}SR spectra for Mo$_2$C. 
The enhanced muon-spin relaxation in the superconducting state is clearly visible
and it is due to the formation of a FLL during the field-cooling process, which generates an inhomogeneous field distribution~\cite{Yaouanc2011,Amato2024,Blundell2021}.
The broadening of field distribution in the superconducting state is clearly reflected in the fast Fourier transform (FFT) 
of the TF-{\textmu}SR spectra [see Figs.~\ref{fig:TF_muSR}(b)-(c)]. 
To describe the field distribution, the TF-{\textmu}SR spectra can be modelled using~\cite{Maisuradze2009}: 
\begin{equation}
	\label{eq:TF_muSR}
	A_\mathrm{TF}(t) = \sum\limits_{i=1}^n A_i \cos(\gamma_{\mu} B_i t + \phi) e^{- \sigma_i^2 t^2/2} +
	A_\mathrm{bg} \cos(\gamma_{\mu} B_\mathrm{bg} t + \phi).
\end{equation}
Here $A_{i}$, $A_\mathrm{bg}$ and $B_{i}$, $B_\mathrm{bg}$ 
are the initial asymmetries and local fields sensed by implanted muons in the 
sample and sample holder,
$\gamma_{\mu}$/2$\pi$ = 135.53\,MHz/T 
is the muon gyromagnetic ratio, $\phi$ is a shared initial phase, and $\sigma_{i}$ 
is the Gaussian relaxation rate of the $i$th component. 
Here, we find that Eq.~\eqref{eq:TF_muSR} with $n = 2$ [solid line in Fig.~\ref{fig:TF_muSR}(b)] shows a better agreement with the experimental data than with $n$ = 1 [dashed line in Fig.~\ref{fig:TF_muSR}(b)]. 
In the normal state, the derived muon-spin relaxation rates $\sigma_i(T)$ are small and independent of temperature while, below $T_c$, they start to increase due to the onset of the FLL
and the increased superfluid density (see inset in Fig.~\ref{fig:rho_sc}).
The effective Gaussian relaxation rate $\sigma_\mathrm{eff}$ can be calculated from $\sigma_\mathrm{eff}^2/\gamma_\mu^2 = \sum_{i=1}^2 A_i [\sigma_i^2/\gamma_{\mu}^2 - \left(B_i - \langle B \rangle\right)^2]/A_\mathrm{tot}$~\cite{Maisuradze2009}, where $\langle B \rangle = (A_1\,B_1 + A_2\,B_2)/A_\mathrm{tot}$ and $A_\mathrm{tot} = A_1 + A_2$. By considering the constant nuclear
relaxation rate $\sigma_\mathrm{n}$ in the narrow temperature
range ($\sim$0.3--5\,K) investigated here, confirmed also 
by ZF-{\textmu}SR measurements (see Fig.~\ref{fig:ZF_muSR}), the
superconducting Gaussian relaxation rate can be extracted from
$\sigma_\mathrm{sc} = \sqrt{\sigma_\mathrm{eff}^{2} - \sigma^{2}_\mathrm{n}}$. 
%
%=== begin figure ==========================%
%=== end figure ==========================%

The effective magnetic penetration depth $\lambda_\mathrm{eff}$ can 
then be calculated using $\sigma_\mathrm{sc}^2(T)/\gamma^2_{\mu} = 0.00371\Phi_0^2/\lambda_\mathrm{eff}^4(T)$~\cite{Barford1988,Brandt2003}.
Figure~\ref{fig:rho_sc} summarizes the temperature-dependent inverse square of magnetic penetration depth $\lambda_\mathrm{eff}^{-2}(T)$, which is proportional to the superfluid density $\rho_\mathrm{sc}(T)$. 
The various models used to analyze the $\rho_\mathrm{sc}(T)$ data, are generally described by the relation:\\ 
\begin{equation}
	\label{eq:rhos}
	\rho_\mathrm{sc}(T) = 1 + 2\, \Bigg{\langle} \int^{\infty}_{\Delta_\mathrm{k}} \frac{E}{\sqrt{E^2-\Delta_\mathrm{k}^2}} \frac{\partial f}{\partial E} \mathrm{d}E \Bigg{\rangle}_\mathrm{FS}. 
\end{equation}
Here, $f = (1+e^{E/k_\mathrm{B}T})^{-1}$ is the Fermi function and $\langle \rangle_\mathrm{FS}$ represents an average over the Fermi surface~\cite{Tinkham1996}; $\Delta_\mathrm{k}(T) = \Delta(T) \delta_\mathrm{k}$ is an angle-dependent
gap function, where $\Delta$ is the maximum gap value and $\delta_\mathrm{k}$ is the 
angular dependence of the gap, equal to 1, $\cos2\phi$, and $\sin\theta$ 
for an $s$-, $d$-, and $p$-wave model, respectively, where $\phi$ 
and $\theta$ are the azimuthal angles. The temperature-dependent gap is assumed to follow $\Delta(T) = \Delta_0 \mathrm{tanh} \{1.82[1.018(T_\mathrm{c}/T-1)]^{0.51} \}$~\cite{Tinkham1996,Carrington2003}, where $\Delta_0$ is the zero-temperature gap value.
The $s$- and $p$-wave models (see black solid and red dashed lines in Fig.~\ref{fig:rho_sc}) yield the same zero-temperature magnetic penetration depth $\lambda_\mathrm{0} =291(3)$\,nm, but different zero-temperature energy gaps $\Delta_0$ =  0.44(1) and 0.60(1)\,meV, respectively. The magnetic penetration depth of the $\beta$-phase Mo$_2$C is much higher than that of the $\alpha$-phase MoC$_x$ ($\lambda_0$ $\sim$ 132\,nm) and the $\eta$-phase Mo$_3$C$_2$ ($\lambda_0$ $\sim$ 197\,nm)~\cite{Sathish2012,Yamaura2006}.
A possible $d$-wave model (green dash-dotted line in Fig.~\ref{fig:rho_sc}) provides a gap size $\Delta_0$ = 0.55(1)\,meV. This is comparable to the $p$-wave model, 
but $\lambda_\mathrm{0}$ = 255(3)\,nm is much shorter than that of both the $s$- and $p$-wave models. 
As can be clearly seen in Fig.~\ref{fig:rho_sc}, below $\sim$1.2\,K, the $d$-wave model deviates significantly from the experimental data. 
At the same time, also the $p$-wave model shows a poor agreement with data in the 0.7--1.6\,K range. The $s$-wave model, on the other hand,
reproduces the experimental data quite well over the entire temperature
range studied. In addition, the temperature-independent
$\lambda_\mathrm{eff}^{-2}(T)$ at $T <$ 1\,K (i.e., $1/3T_c$) definitely
excludes possible gap nodes and suggests that a fully-gapped
superconducting state occurs in Mo$_2$C.

%
%==== figure =============================%
%=== end figure ==========================%

\subsection{\label{ssec:ZF_muSR}Zero-field \texorpdfstring{{\textmu}SR}{MuSR}}
ZF-{\textmu}SR is one of the few techniques sensitive enough to detect the tiny 
spontaneous magnetic field occurring below the superconducting transition temperature. 
Similarly, it is suitable also for detecting a possible short-range magnetic order or magnetic fluctuations. In view of this, 
we performed also ZF-{\textmu}SR measurements on Mo$_2$C. The ZF-{\textmu}SR spectra collected in the normal- and superconducting states of Mo$_2$C are presented in Fig.~\ref{fig:ZF_muSR}.
The lack of a fast decay and of coherent oscillations in the ZF-{\textmu}SR data confirms the absence of magnetic order and/or fluctuation in Mo$_2$C. As a consequence, owing to the absence of magnetic 
fields of electronic origin, the muon-spin relaxation is 
mainly due to the randomly oriented nuclear magnetic moments.
Considering that both Mo and C atoms have relatively small nuclear moments (<1~{\textmu}$_\mathrm{n}$), Mo$_2$C exhibits a very
weak muon-spin relaxation. Therefore, the ZF-{\textmu}SR spectra can be modeled by a Lorentzian-type Kubo-Toyabe relaxation function $G_\mathrm{KT} = [\frac{1}{3} + \frac{2}{3}(1 -\Lambda_\mathrm{ZF}t)\,\mathrm{e}^{-\Lambda_\mathrm{ZF}t}]$~\cite{Kubo1967,Yaouanc2011}, where $\Lambda_\mathrm{ZF}$ represents the zero-field Lorentzian relaxation rate. 
As shown by solid lines in Fig.~\ref{fig:ZF_muSR}, the ZF-{\textmu}SR spectra of Mo$_2$C were fitted to $A_\mathrm{ZF}(t) = A_\mathrm{s} G_\mathrm{KT} + A_\mathrm{bg}$, where $A_\mathrm{s}$ is the same as $A_i$ in Eq.~\eqref{eq:TF_muSR}. The obtained muon-spin relaxation rates are $\Lambda_\mathrm{ZF}$ = 0.021(1)\,{\textmu}s$^{-1}$ at 0.3\,K and 0.019(1)\,{\textmu}s$^{-1}$ at 10\,K.
Obviously, the relaxation rates are almost identical in the superconducting- and the normal state of Mo$_2$C, differing less than 
their standard deviations. The absence of an additional muon-spin relaxation below $T_c$ definitely excludes a possible time-reversal symmetry (TRS) breaking in the superconducting state of Mo$_2$C.
Hence, combined with TF-{\textmu}SR data, our ZF-{\textmu}SR 
results suggest a conventional fully-gapped bulk SC with a preserved TRS in the $\beta$-phase Mo$_2$C superconductor.

%==== figure =============================%
\begin{figure}[!htp]
	\centering
	\includegraphics[width = 0.48\textwidth]{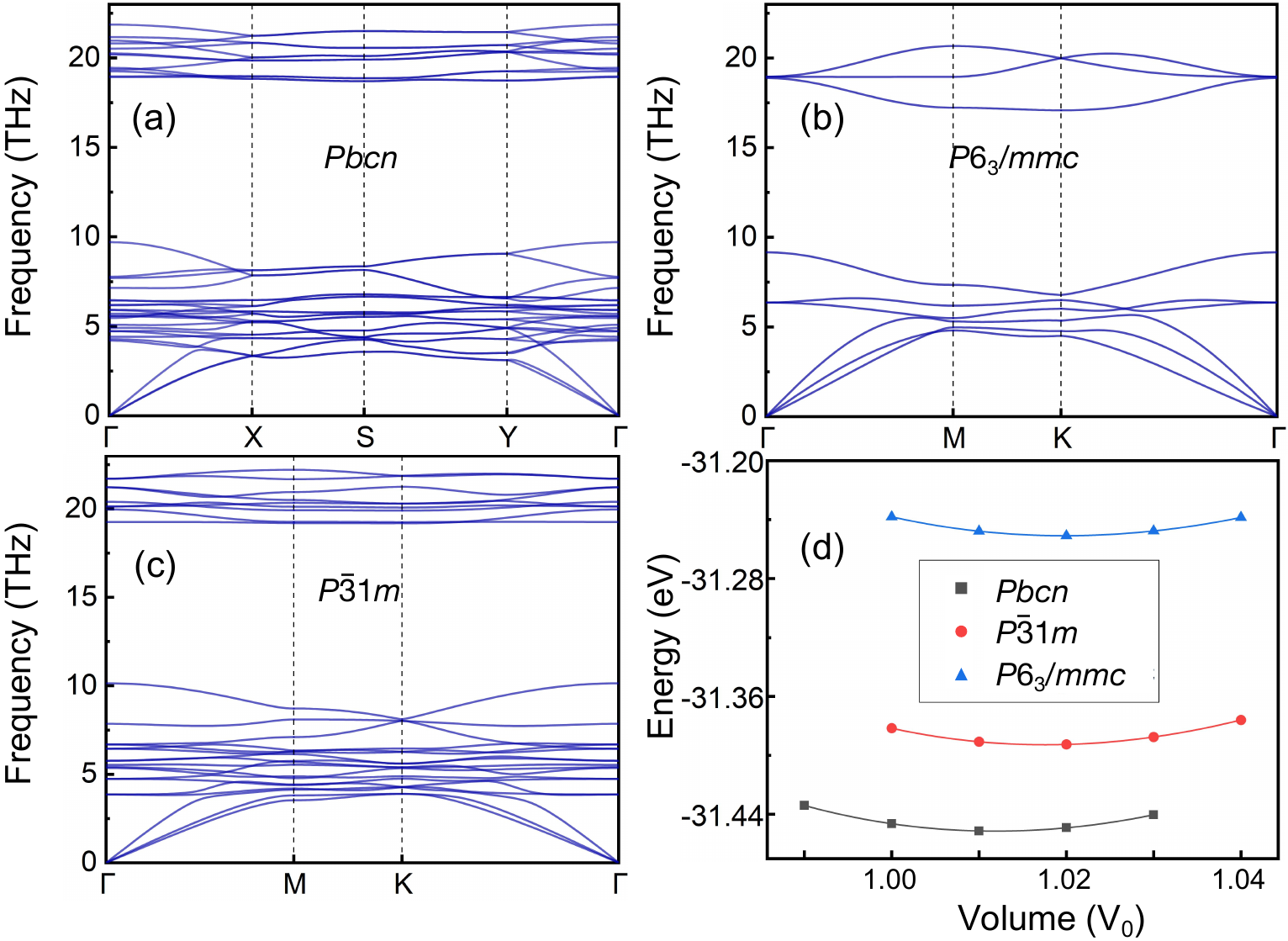}
	%\vspace{-2ex}%
	\caption{The calculated phonon dispersion spectra for the
	$\beta$- (a), $\eta$- (b), and $\zeta$- phase (c) of Mo$_2$C. 
	The calculated total energies vs.\ unit-cell volume for the three crystal structures are shown in (d). Here, $V_0$ stands for the experimental volume (see details in Table~\ref{tab:lattice}).
	%The calculated total energies vs. the measured unit-cell volume $V_0$ for the three crystal structures are shown in (d) (see details in Table~\ref{tab:lattice})
	}	
 	\label{fig:Phonon}
\end{figure}
%=== end figure ==========================%

% 
\subsection{\label{ssec:DFT} Band-structure calculations}  
According to XRD refinements (see Fig.~\ref{fig:XRD}), in Mo$_2$C, the orthorhombic crystal structure shows the best agreement with the 
XRD pattern. To confirm the crystal structure of Mo$_2$C, we performed comparative first-principle calculations of the phonon dispersion spectra of Mo$_2$C by using the space groups $Pbcn$ ($\beta$-phase), $P6_3/mmc$ ($\eta$-phase), and $P\bar{3}1m$ ($\zeta$-phase), respectively. 
As shown in Figs.~\ref{fig:Phonon}(a)-(c), no soft phonon modes could be
identified in the spectra of these structures, implying that all of them are dynamically stable and can be synthesized experimentally. This is consistent with the mixture of different phases
we find in the samples obtained by arc melting. The calculated total energies versus the unit-cell volumes are summarized in Fig.~\ref{fig:Phonon}(d) 
for the three crystal structures. Among them, the $\beta$-phase Mo$_2$C has the lowest energy at the equilibrium volume, while this is highest for the $\eta$-phase. Therefore, the $\beta$-phase molybdenum carbides can be stabilized at a relatively low pressure and temperature compared to the other phases~\cite{Yamaura2006,Sathish2012,Sathish2014,Ge2021}.
Since both the experiment and the theory confirm that Mo$_2$C adopts the $\beta$-phase, we calculated the electronic-band structures solely for this phase. The theoretical results for the other phases can be found elsewhere~\cite{Huang2018,Kavitha2016,Liu2013}.

%==== figure =============================%
\begin{figure}[t]
	\centering
	\includegraphics[width = 0.43\textwidth]{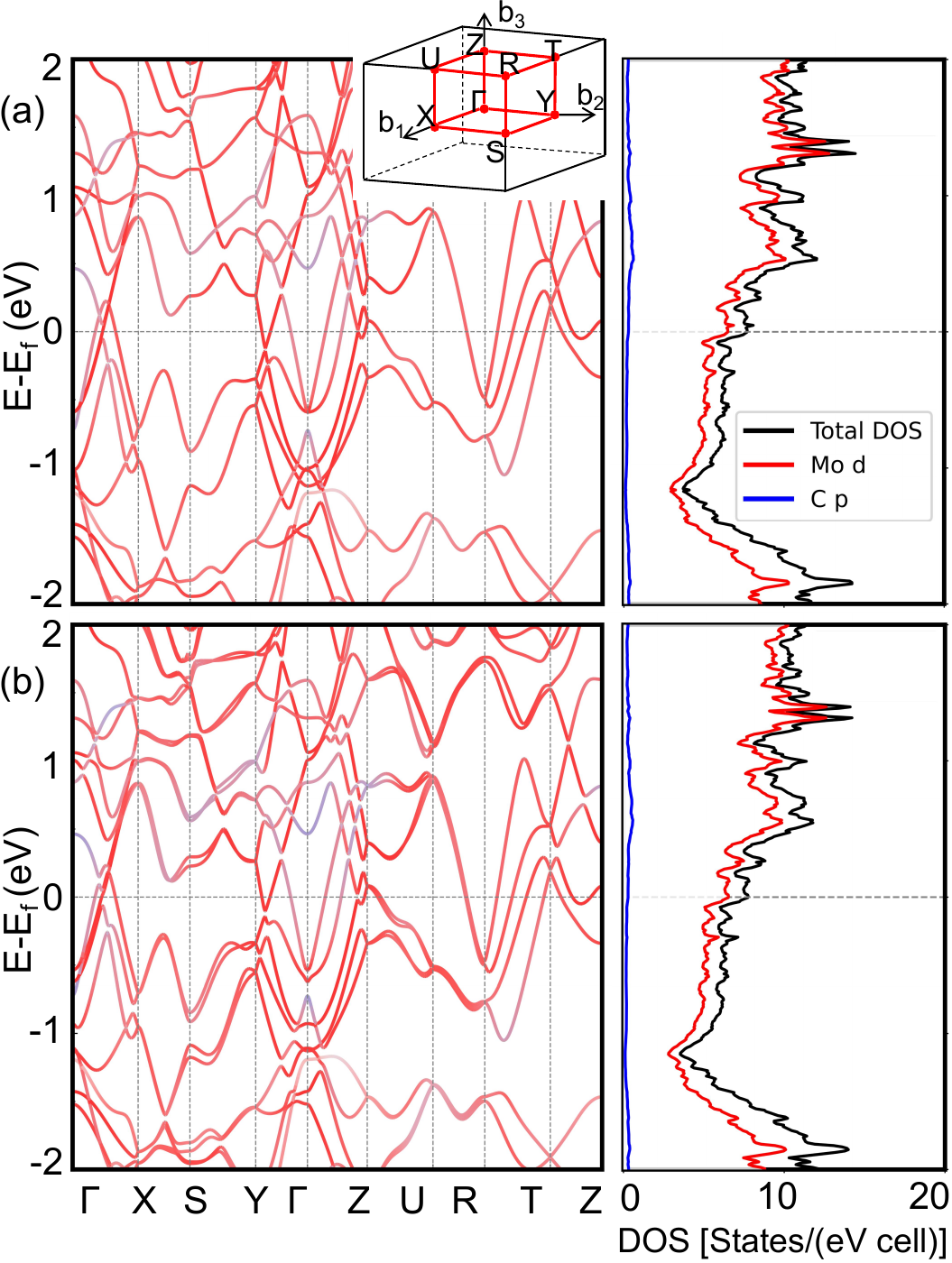}
	%\vspace{-2ex}%
	\caption{Electronic band structures of Mo$_2$C calculated by 
	ignoring (a) and by considering (b) the spin-orbit 
	coupling. Here, the Mo $d$-orbitals and 
	the C $p$-orbitals are shown in red and blue, respectively. The total and partial (Mo and C atoms) density of 
	states without SOC and with SOC are shown on the right side of each panel.
	The primitive-cell Brillouin zone, including the high-symmetry points, is shown in the inset of panel (a).}
\label{fig:Band}
\end{figure}
%=== end figure ==========================%
%

The calculated electronic band structures for the $\beta$-phase 
Mo$_2$C, are summarized in Fig.~\ref{fig:Band}.
Close to the Fermi level, the electronic bands are dominated by the $4d$-orbitals of Mo atoms, while the contribution from the C $2p$-orbitals is almost negligible. Indeed, over a wide range of energies, 
the contribution from the C-$2p$ orbitals is less than 4.4\%. This situation is also reflected in the DOS shown in the right panels.
The estimated DOS at the Fermi level is about 1.93 states/(eV f.u.) [= 7.72 states/(eV cell)/$Z$, with $Z = 4$, the number of Mo$_2$C formula units per unit cell]. Such a relatively high DOS suggests a 
good metallicity for Mo$_2$C, consistent with previous electrical resistivity data~\cite{Ge2021}. After including the SOC the bands separate, since SOC breaks the band degeneracy and brings one of the bands closer to the Fermi level [see Fig.~\ref{fig:Band}(b)]. The band splitting due to the SOC is rather weak, here visible only along the $Z$--$U$ line near the Fermi level. Although the band splitting along the $S$--$Y$ line is quite significant, these bands are too far away from the Fermi level to have any meaningful influence on the electronic properties of Mo$_2$C. In the $\beta$-phase Mo$_2$C, the band splitting $E_\mathrm{SOC}$ is up to 100\,meV. This is comparable to the $\alpha$-phase NbC, but much smaller than in TaC~\cite{Shang2020}. 

%==== figure =============================%
\begin{figure}[t]
	\centering
	\includegraphics[width = 0.42\textwidth]{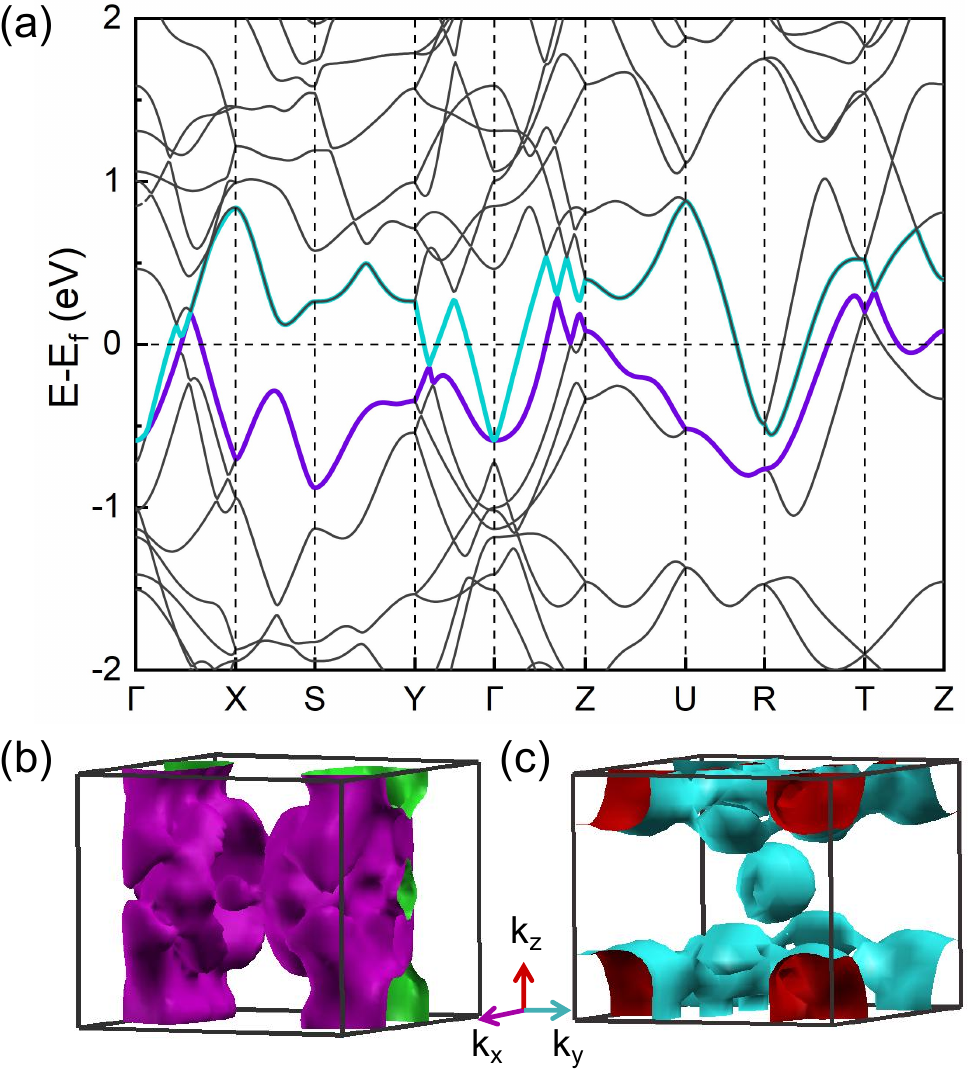}
	%\vspace{-2ex}%
	\caption{(a) Close-up view of the electronic band structure without SOC for Mo$_2$C. The bands that cross the Fermi level are highlighted in purple and cyan. 
		(b)-(c) Representative Fermi surfaces of Mo$_2$C using the same color code for the bands as shown in panel (a).}
\label{fig:FS_Mo2C}
\end{figure}
%=== end figure ==========================%
%
	
The $Pbcn$ space group of Mo$_2$C is nonsymmorphic and it has an inversion symmetry. 
After inspecting the band structure without SOC across the whole Brillouin zone [see Fig.~\ref{fig:Band}(a)], the bands along the $X$--$S$--$Y$, 
$Z$--$U$, and $R$--$T$--$Z$ directions turn out to be 
twofold degenerate, while the bands along $U$--$R$ are fourfold 
degenerate (without considering the spin degree of freedom).
By using symmetry arguments~\cite{yu_encyclopedia_2022}, the $X$--$U$--$S$--$R$ and $U$--$R$--$Z$--$T$ planes are twofold degenerate nodal surfaces due to the combined 
presence of a screw rotation and time-reversal symmetries. 
The fourfold degenerate nodal lines along $U$--$R$ are protected 
by the combination of glide-mirror- and \emph{PT} symmetries.
In the presence of SOC, the fourfold-degenerate bands are broken into two twofold-degenerate bands [see Fig.~\ref{fig:Band}(b)], 
except for the fourfold-degenerate nodal lines along the 
$R$--$T$--$Z$ direction, which are protected by a combination of glide-mirror- and \emph{PT} symmetries.
Therefore, similar to other phases of carbides~\cite{Shang2020,Huang2018}, $\beta$-Mo$_2$C with nodal lines crossing the Fermi level could also be material candidates for future studies of topological superconductivity. 

Among the 8 bands crossing the Fermi level, only two of them 
contribute significantly to the DOS and have the largest Fermi surfaces (FSs). These two bands are highlighted in purple and cyan 
in Fig.~\ref{fig:FS_Mo2C}(a), and their corresponding FSs are depicted in Figs.~\ref{fig:FS_Mo2C}(b) and (c), respectively. Clearly, these two bands form distinct FSs, 
even though both are due to Mo 4$d$-orbitals. 
The purple band exhibits two small hole pockets near the Brillouin center, which are much smaller than the analogous electron 
pocket of the cyan band. Near the Brillouin boundary of the purple band two cylinder-like FSs extend along the $\Gamma$--$Z$ direction. By contrast, in the cyan band, such FSs extend along the $\Gamma$--$Y$ direction. 
Clearly, the FSs of the orthorhombic Mo$_2$C are more three dimensional and more complex than those of the $\alpha$-phase TMCs.
In the latter case, the largest FSs consist of three cylinders along the $k_i$ ($i = x, y, z$) directions.
Such cylinder-like FSs originate from the strong hybridization between the transition metal $d$-orbitals and C $p$-orbitals. 
By contrast, the $p$--$d$ hybridization is rather weak in the orthorhombic Mo$_2$C.  
The cylinderlike FSs are known to play an important role in the SC of high-$T_c$ iron-based materials~\cite{Mazin2008,Kuroki2009,Mazin2019}. This may also be the case for $\alpha$-phase TMCs, which have relatively
high $T_c$ values in comparison to other carbide phases.

\section{Discussion}
Now, we briefly discuss the different phases of molybdenum carbides. 
To date, there are mainly two phases of molybdenum carbides that have been reported to become superconducting at low temperature, namely $\alpha$-MoC$_{x}$ and $\eta$-Mo$_3$C$_2$. The $\gamma$-MoC and $\zeta$-phase Mo$_2$C adopt a noncentrosymmetric hexagonal and a centrosymmetric trigonal structure, respectively, but no SC has been reported in these phases yet. Recent theoretical work predicts that by introducing hole carriers, the $\gamma$-phase MoC could show SC with a $T_c$ up to 9\,K~\cite{Huang2018}. Here, by using the {\textmu}SR technique, we reveal
that it is the $\beta$-phase Mo$_2$C, instead, to represent the third
member of molybdenum carbides to show bulk SC. 
Among the latter, the $\alpha$- and $\beta$-phases show the highest and the lowest $T_c$, i.e., $\sim 15$\,K~\cite{Sathish2012,Sathish2014} and $\sim 3.2$\,K, respectively. While the $\eta$-Mo$_3$C$_2$-phase shows an intermediate $T_c$ of 7.4\,K~\cite{Yamaura2006}.
The highest $T_c$ in the $\alpha$-phase TMCs is most likely due to their
strong $p$--$d$ hybridization and, thus, to an enhanced electron-phonon coupling.
We recall that, the strong $p$--$d$ hybridization produces large
cylinder-like FSs~\cite{Shang2020,Huang2018}, which play an important
role also in the SC of high-$T_c$ iron-based materials~\cite{Mazin2008,Kuroki2009,Mazin2019}. 
As for the $\beta$-phase Mo$_2$C, band-structure calculations indicate
a rather weak $p$--$d$ hybridization (see Fig.~\ref{fig:Band}), which
may justify their comparatively low $T_c$ value. 

The low-temperature superfluid density, determined by TF-{\textmu}SR
in our study, suggests a fully-gapped superconducting state in the $\beta$-phase Mo$_2$C. A {\textmu}SR study has not yet been performed in the $\alpha$-phase MoC$_x$ and $\eta$-phase Mo$_3$C$_2$. 
This is related to the difficulties in synthesizing sufficient amounts of material under the demanding conditions (1700\,$^\circ$C, 6--17\,GPa) required in these cases~\cite{Sathish2012,Sathish2014,Yamaura2006}. 
According to our previous TF-{\textmu}SR studies, the $\alpha$-phase NbC and TaC also exhibit a fully-gapped superconducting state~\cite{Shang2020}. We expect also the $\alpha$-phase MoC$_x$ to show similar SC properties to NbC and TaC.
In fact, the electronic specific heat of $\alpha$-phase MoC$_x$
(and $\eta$-phase Mo$_3$C$_2$) shows an exponential temperature dependence
in the superconducting state, consistent with a nodeless SC~\cite{Sathish2012,Sathish2014,Yamaura2006}. Further, the small
zero-temperature energy gap ($\Delta_0$ < 1.76~$\mathrm{k}_\mathrm{B}$$T_c$)
and a reduced specific-heat jump at $T_c$ ($\Delta$$C$/$\gamma$$T_c$ < 1.43) suggest a weakly coupled SC in the various phases of superconducting molybdenum carbides. Taking into account the preserved TRS in the superconducting state, as well as an upper critical field $H_\mathrm{c2}$ well below the Pauli limit~\cite{Shang2020,Sathish2012,Yamaura2006}, we conclude that the molybdenum
carbides exhibit a spin-singlet pairing, independent of
their crystal structure (phase). 

Finally, we discuss the topological aspects of molybdenum carbides. The $\alpha$-phase MoC, possesses a nonzero $\mathbb{Z}_2$ topological invariant and Dirac surface states~\cite{Huang2018}. 
The isostructural NbC, contains three closed node lines in the bulk band structure (without considering SOC) of its first Brillouin zone. These are protected by time-reversal and space-inversion symmetry~\cite{Shang2020}.
In case of a large SOC, such nodal loops become gapped. Since the $4d$ Nb and
Mo atoms exhibit a weaker intrinsic SOC than the $5d$ Ta atoms,
the SOC effects should be modest in both NbC and MoC. Consequently, the node lines --- predicted by calculations neglecting SOC effects --- are most likely preserved in  both the above carbides. As such, the $\alpha$-phase MoC and NbC might 
be good candidates for observing the exotic two-dimensional surface states. Further, although the $\gamma$-phase MoC is not superconducting in its pristine form, it is predicted to be a topological nodal-line material, exhibiting drumhead surface states. 
After introducing hole carriers, its SC can be tuned to reach a $T_c$ of up to 9\,K~\cite{Huang2018}.  
Since the $\gamma$-phase MoC adopts a noncentrosymmetric hexagonal structure, it can be classified as a topological Weyl semimetal. Indeed, three-component fermions were experimentally observed in the $\gamma$-phase MoP and WC~\cite{Lv2017,Ma2018}. By applying external pressure, the topological semimetal MoP becomes a superconductor, whose $T_c$ reaches 4\,K (above 90\,GPa)~\cite{Chi2018}, thus representing 
a possible candidate topological superconductor. Here, we also find that the $\beta$-phase Mo$_2$C hosts twofold-degenerate nodal surfaces and fourfold-degenerate nodal lines near the Fermi level. In the case of SOC, the fourfold degenerate
nodal lines cross the Fermi level and, hence, could contribute to the
superconducting pairing. In general, all the various phases of molybdenum carbides are promising for studying topological superconductivity.

\vspace{7pt}
\section{\label{ssec:Sum}Conclusion}

To summarize, we studied the superconducting properties of %the 
Mo$_2$C mostly by means of the {\textmu}SR technique, 
as well as via numerical band-structure calculations.
The latter show that the phonon dispersion spectrum of Mo$_2$C
provides the lowest total energy in case of the orthorhombic $\beta$-phase
(with $Pbcn$ space group), a result consistent with the experiment.
Magnetization measurements confirm the bulk superconductivity
of Mo$_2$C, with a $T_c$ of 3.2\,K. The temperature dependence of the superfluid density reveals a \emph{nodeless} 
superconducting state, which is well described by an \emph{isotropic $s$-wave}
model. The lack of spontaneous magnetic fields below $T_c$ indicates 
that time-reversal symmetry is \emph{preserved} in the superconducting state of Mo$_2$C. 
Electronic band-structure calculations suggest that the density of states
at the Fermi level is dominated by the Mo-$4d$ electrons, while the
contribution of the C-$2p$ electrons is negligible over a broad
energy range. As a consequence, the $p$--$d$ hybridization is rather weak
in the $\beta$-phase Mo$_2$C, resulting in a relatively low
$T_c$ value. Topological nodal states including nodal surfaces and nodal lines could be identified
in the Mo$_2$C electronic band structure near the Fermi level.
This finding, together with the intrinsic superconductivity,
suggests that the $\beta$-phase Mo$_2$C, too, is a potential candidate for %future
studies of topological SC, similar to the other phases of molybdenum carbides.

\begin{acknowledgments} 
The authors thank Weikang Wu for fruitful discussions.
This work was supported by the Natural Science Foundation of Shanghai
(Grant Nos.\ 21ZR1420500 and 21JC\-140\-2300), Natural Science
Foundation of Chongqing (Grant No.\ CSTB-2022NSCQ-MSX1678), National
Natural Science Foundation of China (Grant No.\ 12374105), Fundamental
Research Funds for the Central Universities, and the Schwei\-ze\-rische
Na\-ti\-o\-nal\-fonds zur F\"{o}r\-der\-ung der Wis\-sen\-schaft\-lichen
For\-schung (SNF) (Grant Nos.\ 200021\_169455 and No.\ 200021\_188706). 
We also acknowledge the allocation of beam time at the Swiss muon source (Dolly {\textmu}SR spectrometer).
\end{acknowledgments}

%\begin{footnotesize}
\bibliography{Mo2C}
%\end{footnotesize}

\end{document}